\documentclass{IEEE_lsens}

\usepackage[font=scriptsize,caption=false,labelsep=space]{subfig}
\usepackage{mathrsfs}
\usepackage{graphicx,float,wrapfig,epstopdf,amsmath}
\usepackage[]{algorithmicx}
\usepackage{algpseudocode,algorithm}
\usepackage{balance}

\usepackage{amsmath,mathtools }
\usepackage{amssymb}
\usepackage{amsthm}
\usepackage{graphicx}
\usepackage{epstopdf}
\usepackage{times}
\usepackage{textcomp,cite}
\usepackage{color}
\usepackage{url}
\usepackage{cuted}

\usepackage{multirow}
\usepackage[table]{xcolor}
\usepackage{threeparttable}

\begin{document}
	
	\title{DeepMUSIC: Multiple Signal Classification via Deep Learning
	}
	\author{Ahmet~M.~Elbir\textit{, Senior Member, IEEE} 
		\thanks{A. M. E. is with the Department of Electrical and Electronics Engineering, Duzce University, Duzce, Turkey. e-mail: ahmetelbir@duzce.edu.tr, ahmetmelbir@gmail.com.}
		
	}
	\maketitle
	
	\begin{abstract}
		This letter introduces a deep learning (DL) framework for the classification of multiple signals in direction finding (DF) scenario via sensor arrays. Previous works in DL context mostly consider a single or two target scenario which is a strong limitation in practice. Hence, in this work, we propose a DL framework called DeepMUSIC for multiple signal classification. We design multiple deep convolutional neural networks (CNNs), each of which is dedicated to a subregion of the angular spectrum. Each CNN learns the MUSIC spectra of the corresponding angular subregion. Hence, it constructs a non-linear relationship between the received sensor data and the angular spectrum. We have shown, through simulations, that the proposed DeepMUSIC framework has superior estimation accuracy and exhibits less computational complexity in comparison with both DL and non-DL based techniques.
	\end{abstract}

	\section{Introduction}
	\label{sec:Introduciton}
	Direction finding (DF) is a crucial task for direction-of-arrival (DOA) estimation in a variety of fields including, radar, sonar, acoustics and communications~\cite{krim_viberg}. While there are several different approaches in the literature, the MUSIC (MUltiple SIgnal Classification) algorithm~\cite{music} is the most popular method for this purpose. 
	
	In the literature, most of the algorithms are model-based approaches such that the performance of the DOA estimation algorithms strongly relies on the perfectness of the input data~\cite{friedlander}. In order to mitigate this drawback, learning-based and data-driven architectures are proposed so that the non-linear relationship between the input and output data can be learned by neural networks~\cite{deepDOAEstMLP_DoaEstimation,deepDOAEstSparsePrior,deepDOAestUCA}. Hence, as a class of machine learning, deep learning (DL) has gained much interest recently. DL is capable of uncovering complex relationships in data/signals and, thus, can achieve better performance. {\color{black}While  there are several papers to demonstrate the performance of DL in wireless communications~\cite{deepLearning_FastBeamforming,deepLearning_ModulationClassification,elbirQuantizedCNN2019}, limited number of works are considered in the context of DOA estimation and array signal processing~\cite{deepDOAestUCA,deepDOAreviewDOAEstimation}.} 
	
	DOA estimation via DL is considered in~\cite{deepDOAEstMLP_DoaEstimation} where a multilayer perceptron (MLP) architecture is proposed to resolve two target signals. In \cite{deepDOAEstSparsePrior}, the authors studies the same problem, also for two signal case, by exploiting the sparsity of the received signal in angular domain and design a deep convolutional neural network (CNN). A single sound target case is assumed in \cite{deepDOA_widebandSingleSoundSource} and an MLP architecture is proposed to estimate the target DOA angle for wideband case. Acoustic scenario is also studied in~\cite{deepDOA_onlineSingleSource} by incorporating long short term memory (LSTM) with CNN for online DOA estimation, which is also performed for a single target case. In a recent work \cite{elbirIETRSN2019},  cognitive radar scenario is considered where DL is applied for sparse array selection and DOA estimation for a single target. This approach is extended for two targets in~\cite{elbirSampTA} for sparse array design.
	
	A common assumption in above works is that the number of targets is assumed to be small. This is because the complexity of the generation of the training data and the training overhead become more difficult as the number of targets increases. Specifically, the data set length increases on the order of $N^K$ for $K$ being the number of targets and $N$ is the number grid points in the angular spectrum. In order to reduce the complexity, in this work, we proposed a multiple deep network approach for multiple target estimation. In particular, we design multiple CNNs, each of which is dedicated for a subregion of the angular spectrum. Hence, we partition the angular spectrum into non-overlapping subregions, and assume that there is a single target in each subregion. This assumption is relevant \cite{deepDOAEstMLP_DoaEstimation,deepDOAestUCA} and it can be generalized for higher number of targets with close separation by simply increasing the number of deep networks. In order to feed the deep networks, the covariance of the sensor outputs is used as a common input. Then, the output label of each network is the MUSIC spectra of the corresponding angular subregion. Hence, we call the proposed approach \textsf{DeepMUSIC} which yields the MUSIC spectra at the output. {\color{black} The main contributions of the proposed DL approach are as follows. 1) We have introduced a DL approach to estimate multiple target DOAs whereas the previous works can only work for limited number of targets. 2)\textsf{DeepMUSIC} provides less computation time as compared to both DL- and non-DL-based approaches.  2) \textsf{DeepMUSIC}  has higher DOA estimation accuracy than the conventional DL-based techniques. It provides asymptotic performance for moderate SNR levels and it performance maxes out in high SNR due to the loss of precision because of the biased nature of DL-based approaches.}

	\section{Array Signal Model}
	Consider $K$ far-field signals impinging on an $M$-element uniform linear array (ULA). Then, the output of the antenna array in the baseband can be given by 
	\begin{align}
	\label{arrayOutput}
	\mathbf{y}(t_i) = \sum_{k=1}^{K} \mathbf{a}(\theta_k) s_k(t_i) + \mathbf{n}(t_i),\hspace{5pt} i = 1,\dots,T,
	\end{align}
	where $T$ is the number of data snapshots and ${s}_k(t_i)\in \mathbb{C}$ is the signal emitted from the $k$-th target which is located with the DOA angle $\theta_k $ with respect to the antenna array. $\mathbf{a}(\theta_k)\in \mathbb{C}^{M}$ denotes the array steering vector whose $m$-th element is given by
	\begin{align}
	a_m(\theta_k) = \exp\{-j \frac{2\pi \bar{ d}(m-1)}{\lambda} \sin(\theta_k)  \},
	\end{align}
	where $\lambda=\frac{c_0}{f_c}$ is the wavelength for $f_c$ being the carrier frequency and $c_0$ is the speed of light. $\mathbf{n}(t_i) \sim \mathcal{N}(\mathbf{0}_M,\sigma_n^2\mathbf{I}_M)$ is zero-mean spatially and temporarily white additive Gaussian noise vector which corrupts the emitted signal with variance $\sigma_n^2$. Using the array output in (\ref{arrayOutput}) the covariance matrix of the received signal can be written as
	\begin{align}
	\label{cov1}
	\mathbf{R}_\mathrm{y} = \mathbb{E}\{ \mathbf{y} \mathbf{y}^\textsf{H}  \} = \mathbf{A} \boldsymbol{\Gamma} \mathbf{A}^\textsf{H} + \sigma_n^2 \mathbf{I}_M,
	\end{align}
	where $\mathbb{E}\{\cdot\}$ denotes the expectation operation, $\boldsymbol{\Gamma} = \mathrm{diag}\{\sigma_1^2,\sigma_2^2,\dots, \sigma_K^2\}$ is a $K\times K$ matrix whose diagonal entries are the signal variances and $\mathbf{A}$ is the array steering matrix defined as $\mathbf{A} = [\mathbf{a}(\theta_1),\mathbf{a}(\theta_2),\dots,\mathbf{a}(\theta_K)]\in\mathbb{C}^{M\times K}$\footnote{While (\ref{cov1}) requires the uncorrelated signal assumption, the proposed DL approach can also work well for correlated/coherent signals since the non-linear mapping provided by \textsf{DeepMUSIC} does not rely on the statistical properties of the signal~\cite{elbirIETRSN2019}. To generate the output label for a coherent scenario, the MUSIC spectra can be obtained by employing  spatial smoothing algorithm.}.  Through eigendecomposition, we can rewrite (\ref{cov1}) as $	\mathbf{R}_\mathrm{y} = \mathbf{U}\boldsymbol{\Lambda} \mathbf{U}^\textsf{H},$
	where $\boldsymbol{\Lambda}$ is a diagonal matrix composed of the eigenvalues of $\mathbf{R}_\mathrm{y}$ in descending order as $\boldsymbol{\Lambda} = \mathrm{diag}\{\lambda_1,\lambda_2,\dots,\lambda_M\}$ and $\mathbf{U} = [\mathbf{U}_\mathrm{S}\; \mathbf{U}_\mathrm{N}]  $ is an ${M\times M}$ eigenvector matrix whose first $K$ column vectors correspond to the signal subspace  by $\mathbf{U}_\mathrm{S}$ and the remaining $M-K$ column vectors are the noise subspace eigenvectors as $\mathbf{U}_\mathrm{N}\in \mathbb{C}^{M\times M-K}$. Using the orthogonality of signal and noise subspaces (i.e., $\mathbf{U}_\mathrm{S} \perp \mathbf{U}_\mathrm{N} $), and the fact that the columns of $\mathbf{U}_\mathrm{S}$ and $\mathbf{A}$ span the same space, we have $ ||  \mathbf{U}_\mathrm{N}^\textsf{H} \mathbf{A} ||_\mathcal{F} = 0 $ where $\mathcal{F}$ denotes the Frobenious norm~\cite{music}. Now, we can write the MUSIC spectra $P(\theta)$ as
	\begin{align}
	\label{MUSICSpectra}
	P(\theta) = \frac{1}{\mathbf{a}^\textsf{H}(\theta) \mathbf{U}_\mathrm{N}\mathbf{U}_\mathrm{N}^\textsf{H} \mathbf{a}(\theta)   },
	\end{align}
	whose largest  $K$ peaks correspond to the target DOA angles $\{\theta_k\}_{k=1}^K$. To obtain (\ref{MUSICSpectra}) in practice, we use the sample covariance matrix $\hat{\mathbf{R}}_\mathrm{y}$ since $\mathbf{R}_\mathrm{y}$ is not available. As a result,   $\hat{\mathbf{R}}_\mathrm{y} = \frac{1}{T} \sum_{i = 1}^{T} \mathbf{y}(t_i)\mathbf{y}^\textsf{H}(t_i)$ is the input to the deep network.

	%
	
\begin{figure}[t]
		\centering
		{\includegraphics[draft=false,width=.35\textheight,height=.13\textheight]{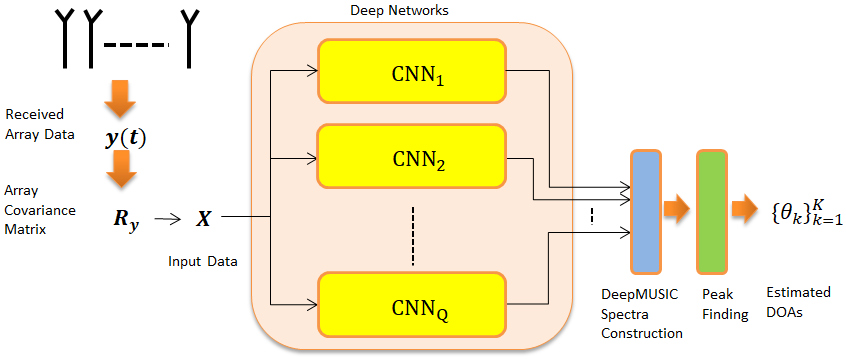} } 
		\caption{The overall \textsf{DeepMUSIC} framework for DOA estimation.}
		\label{fig_systemDiagram}
	\end{figure}

	In this work we can formulate the problem as estimating the target DOA angles $\{\theta_k\}_{k=1}^K$ when the array output $\{\mathbf{y}(t_i)\}_{i=1}^{T}$ is given. For this purpose we introduce a DL framework  as shown in Fig.~\ref{fig_systemDiagram} which is fed by the array covariance matrix $\mathbf{R}_\mathrm{y}$ and it gives the MUSIC spectra $P(\theta)$ at the output.

	
	\section{DOA Estimation via Deep Learning}
	\label{sec:DL}
	The proposed \textsf{DeepMUSIC} framework accepts the array covariance matrix as input and yields the MUSIC spectra at the output. In the following, we first design the labels and the input of the proposed deep networks, then discuss the network architectures and the training.

	In the proposed DL framework shown in Fig~\ref{fig_systemDiagram}, we design $Q$ ($\geq K$) deep networks, each of which is dedicated to a subregion of the angular spectrum. Partitioning the angular spectrum allows us to estimate the multiple target locations more effectively. The use of a single deep network is computationally prohibitive due to the fact that the training data must contain all candidate multiple target locations, whose complexity increase on the order of $N^K$ for $N$ DOA grid points. DOA estimation via partitioned angular spectrum is more efficient such that a reasonable assumption is made such that there is a single target present in each angular subregion \cite{deepDOAEstMLP_DoaEstimation}.
	
	Let $\Theta = \{\Theta^\mathrm{start},\dots,\Theta^\mathrm{final}\}$ be the set of DOA angles where the MUSIC cost function in (\ref{MUSICSpectra}) is evaluated for the starting and final DOA angles $\Theta^\mathrm{start}$ and $\Theta^\mathrm{final}$ respectively.  Then, we denote the MUSIC spectra in (\ref{MUSICSpectra}) by $\mathbf{p}\in \mathbb{R}^{N}$ as $ \mathbf{p}= [P(\Theta^\mathrm{start}),\dots, P(\Theta^\mathrm{final})]^\textsf{T}.$ To obtain the MUSIC spectra for each subregion, we partition $\mathbf{p}$ and $\Theta$ into $Q$ non-overlapping subregions such that $	\Theta = \cup_{q=1}^Q \Theta_q,$
	where each angular set is defined by  
	\begin{align}
	\Theta_q =\{\Theta_q^\mathrm{start},\Theta_q^\mathrm{start} + \gamma, \Theta_q^\mathrm{start} + 2\gamma,\dots,\Theta_q^\mathrm{final}-\gamma\},
	\end{align} 
	where $\gamma = \frac{|\Theta^\mathrm{start}  - \Theta^\mathrm{final} |}{N}$ is the angular resolution and $\Theta_q^\mathrm{final}  = \Theta_{q+1}^\mathrm{start}$. Hence, the number of elements in $\Theta_q$ is  $| \Theta_q|= L$ where $L = N/Q$ which is assumed to be an integer number without loss of generality. We can also  rewrite $\mathbf{p}$  as 
	\begin{align}
	\mathbf{p} =[\mathbf{p}_1^\textsf{T},\mathbf{p}_2^\textsf{T},\dots,\mathbf{p}_Q^\textsf{T}]^\textsf{T}.
	\end{align} In particular, the MUSIC spectra for the $q$-th subregion is represented by an $L \times 1$ real-valued vector as 
	\begin{align}
	\mathbf{p}_q = [P(\Theta_q^\mathrm{start}), P(\Theta_q^\mathrm{start} + \gamma), 
	P(\Theta_q^\mathrm{start} + 2\gamma),\dots, P(\Theta_q^\mathrm{final}-\gamma) ]^\textsf{T}.
	\end{align}
	
	During training, once $\mathbf{p}$ is obtained, we assign the MUSIC spectra of each subregion $\{\mathbf{p}_q\}_{q=1}^Q$ as the labels of each deep network. 	To construct the input data, we use the real, imaginary and the angular values of the covariance matrix $\mathbf{R}_\mathrm{y}$. Let $\mathbf{X}$ be an $M\times M \times 3$ real-valued matrix, then the $(i,j)$-th entry of the first and the second "channels" of $\mathbf{X}$ are given by $[[\mathbf{X}]_{:,:,1}]_{i,j} = \operatorname{Re}\{[\mathbf{R}_\mathrm{y}]_{i,j}\}$ and $[[\mathbf{X}]_{:,:,2}]_{i,j} = \operatorname{Im}\{[\mathbf{R}_\mathrm{y}]_{i,j}\}$ respectively. Similarly, the $(i,j)$-th entry of the third "channel" of $\mathbf{X}$ is defined as $[[\mathbf{X}]_{:,:,3}]_{i,j} =\angle \{[\mathbf{R}_\mathrm{y}]_{i,j}\}$ where $\angle\{\cdot\}$ returns the angle information of a complex quantity. {\color{black} While other input structures are possible such as the real and imaginary parts of the upper triangle of the covariance matrix \cite{deepDOAEstMLP_DoaEstimation}, we observed that the above approach provides better feature extraction performance inherit in the input as well as achieving satisfactory mapping accuracy~\cite{elbirIETRSN2019,elbirQuantizedCNN2019}.}

	\begin{algorithm}[h]
		\begin{algorithmic}[1]
			\caption{Training data generation for \textsf{DeepMUSIC}. }
			\Statex {\textbf{Input:} $J_\alpha$, $J_\beta$, $T$,  $M$, $Q$, $K$,  SNR$_{\text{TRAIN}}$}.
			\label{alg:algorithmTraining}
			\Statex {\textbf{Output:} Training data sets $\{ \mathcal{D}_q\}_{q=1}^Q$.}
			\State Generate $J_\alpha$ DOA angle sets $\{\theta_k^{(\alpha)}\}_{k=1}^K$ such that $\theta_k^{(\alpha)} \in [\Theta_k^\mathrm{start}, \Theta_k^\mathrm{final}]$ for $\alpha = 1,\dots, J_\alpha$.
			\State Initialize with $\mu =1$ while the dataset length is $J=J_\alpha J_\beta$.
			\State   \textbf{for}  $1 \leq \alpha \leq J_\alpha $ \textbf{do}
			\State \indent Construct $\mathbf{A}^{(\alpha)} = [\mathbf{a}(\theta_1^{(\alpha)}),\dots, \mathbf{a}(\theta_K^{(\alpha)})]$.
			\State \indent Construct  $\tilde{\mathbf{R}}_\mathrm{y}^{(\alpha)} = \mathbf{A}^{(\alpha)} \boldsymbol{\Gamma}^{(\alpha)} \mathbf{A}^{(\alpha)}$.
			\State \indent Using $\tilde{\mathbf{R}}_\mathrm{y}^{(\alpha)}$, obtain noise subspace   $\mathbf{U}_\mathrm{N}^{(\alpha)}$.
			\State \indent Compute $P^{(\alpha)}(\theta)$ in (\ref{MUSICSpectra}) for $\theta \in [\Theta^\mathrm{start},\Theta^\mathrm{final}]$ and 
			\par \indent construct $\mathbf{p}^{(\alpha)}$ and the partitioned spectra $\{ \mathbf{p}_q^{(\alpha)} \}_{q=1}^Q$.
			\State  \indent \textbf{for}  $1 \leq \beta \leq J_\beta$ \textbf{do}
			\State \indent\indent Generate $s_k^{(\alpha,\beta)}(t_i) \sim   \mathcal{CN}(\mathbf{0}_K,\mathbf{I}_K)$ for $T$ snapshots.
			\State \indent\indent Generate noisy array output \par \indent \indent
			$\mathbf{y}^{(\alpha,\beta)}(t_i) = \sum_{k=1}^{K} \mathbf{a}(\theta_k^{(\alpha)}) s_k^{(\alpha,\beta)}(t_i) + \mathbf{n}^{(\alpha,\beta)}(t_i)$, 
			\par \indent\indent  where $\mathbf{n}^{(\alpha,\beta)} \sim \mathcal{CN}(\mathbf{0}_M,\sigma_{\text{TRAIN}}^2\mathbf{I}_M)$.
			\State \indent\indent Construct sample covariance matrix
			\par \indent\indent $\mathbf{R}_\mathrm{y}^{(\alpha,\beta)} = \frac{1}{T}\sum_{i=1}^{T} \mathbf{y}^{(\alpha,\beta)} (t_i)\mathbf{y}^{(\alpha,\beta)^\textsf{H}} (t_i)$.
			\State \indent\indent Form the input data $\mathbf{X}^{(\mu)}$ as
			\par \indent\indent $[[\mathbf{X}^{(\mu)}]_{:,:,1}]_{i,j} = \operatorname{Re}\{[\mathbf{R}_\mathrm{y}^{(\alpha,\beta)}]_{i,j}\}$.
			\par \indent\indent $[[\mathbf{X}^{(\mu)}]_{:,:,2}]_{i,j} = \operatorname{Im}\{[\mathbf{R}_\mathrm{y}^{(\alpha,\beta)}]_{i,j}\}$.
			\par \indent\indent $[[\mathbf{X}^{(\mu)}]_{:,:,3}]_{i,j} = \angle\{[\mathbf{R}_\mathrm{y}^{(\alpha,\beta)}]_{i,j}\}$.
			\State  \indent\indent Form the output of the $q$-th CNN as $\mathbf{z}_q^{(\mu)} = \mathbf{p}_q^{(\alpha)}$.
			\State \indent\indent  Design input-output pair for $q$-th CNN as $(\mathbf{X}^{(\mu)},\mathbf{z}_q^{(\mu)} )$.
			\State \indent\indent $\mu = \mu+1$.
			\State \indent\textbf{end for} $\beta$,	
			\State \textbf{end for} $\alpha$,
			\State Training data for the $q$-th CNN is obtained from the collection of the input-output pairs as \par \noindent $\mathcal{D}_q = \big((\mathbf{X}^{(1)}, \mathbf{z}_q^{(1)}),(\mathbf{X}^{(2)}, \mathbf{z}_q^{(2)}),\dots, (\mathbf{X}^{(J)}, \mathbf{z}_q^{(J)})\big).$
		\end{algorithmic}
	\end{algorithm}

\begin{figure}[]
		\centering
		{\includegraphics[draft=false,width=.35\textheight]{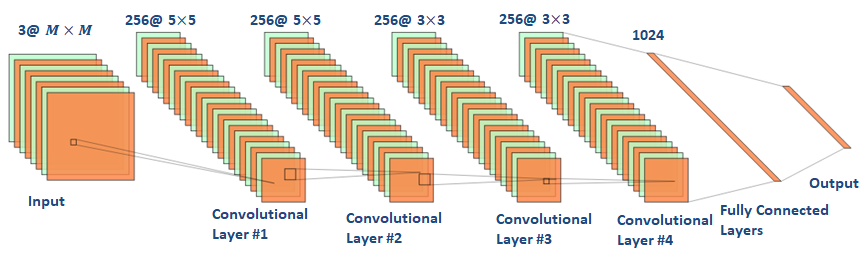} } 
		\caption{Deep network architecture for the proposed algorithm. Each convolutional layer block also includes normalization and $\mathrm{ReLU}$ layers.}
		\label{fig_Network}
	\end{figure}

	We design $Q$ identical CNN structures to estimate the target DOA angles. We demonstrate the network architecture of the proposed CNN structure in Fig.~\ref{fig_Network}. Each CNN is composed of $17$ layers including input and output layers. The overall deep network structure for the $q$-th subregion can be  represented by a non-linear mapping function $\mathbf{\Sigma}_q(\mathbf{X}): \mathbb{R}^{M\times M\times 3}\rightarrow \mathbb{R}^{L}$. In particular, we have 
	\begin{align}
	\label{networkFunction}
	\boldsymbol{\Sigma}_q(\mathbf{X}) =  f^{(17)}\big( f^{(16)} (\cdots   f^{(1)}( \mathbf{X})  \cdots)\big) = \mathbf{p}_q,
	\end{align}
	where $f^{(14)}(\cdot)$ denotes a fully connected layer which maps an arbitrary input $\bar{\bf x}\in \mathbb{R}^{C_x}$ to the  output $\bar{\bf y}\in \mathbb{R}^{C_y}$ by using the weights  $\bar{\bf W} \in \mathbb{R}^{C_{x}\times C_{y}}$. Then the $c_y$-th element of the output of the layer can be given by the inner product
	\begin{align}
	\bar{\bf y}_{c_y} = \langle \bar{\bf W}_{c_y}, \bar{\bf x} \rangle = \sum_{i} {[\bar{\bf W}}]_{c_y,i}^\textsf{T} \bar{ \bf x}_i ,
	\end{align}
	for $c_y = 1,\dots, C_y$ and  $\bar{\bf W}_{c_y}$ is the $c_y$-th column vector of $\bar{\bf W}$ and $C_x = C_y = 1024$ is selected for $f^{(14)}(\cdot)$.
	
	In (\ref{networkFunction}), $\{f^{(i)}(\cdot)\}_{i=\{2, 5, 8, 11\}}$ represent the convolutional layers. The arithmetic operation of a single filter of a  convolutional layer  can be defined for an arbitrary input  $\bar{\bf X} \in \mathbb{R}^{d_{x}\times d_{x}\times V_x}$ and output $\bar{\bf Y} \in \mathbb{R}^{d_{y}\times d_y\times V_{y}}$ as 
	\begin{align}
	\bar{\bf Y}_{p_y,v_y} = \sum_{p_k,p_x} \langle \bar{\bf W}_{v_y,p_k}, \bar{\bf X}_{p_x} \rangle,
	\end{align}
	where   $d_x \times d_y$ is the size of the convolutional kernel, and $V_x \times V_y$ are the size of the response of a convolutional layer. $\bar{\bf W}_{v_y,v_k}\in \mathbb{R}^{V_x}$ denotes the weights of the $v_y$-th convolutional kernel, and $\bar{\bf X}_{p_x} \in \mathbb{R}^{V_x}$ is the input feature map at spatial position $p_x$. Hence we define $p_x$ and $p_k$ as the 2-D spatial positions in the feature maps and convolutional kernels, respectively \cite{quantizedCNN_Unified,deepDOAreviewDOAEstimation}.   In the proposed architecture, we use $256$ filters, the first two of which are of size $5\times 5$ and the remaining two have $3\times 3$ filters.
	
	In (\ref{networkFunction}), $\{f^{(i)}(\cdot)\}_{i=\{3, 6, 9, 12\}}$ are the normalization layers and $\{f^{(i)}(\cdot)\}_{i=\{4, 7, 10, 13\}}$ are the rectified linear unit (ReLU) layers which are defined as $\mathrm{ReLU}(x) = \max(0,x)$. $f^{(15)}(\cdot)$ is a dropout layer and $f^{(16)}(\cdot)$ is a $\mathrm{softmax}$ layer defined for an arbitrary input $\bar{\mathbf{x}}\in \mathbb{R}^{D}$ as $\mathrm{softmax}(\bar{{x}}_i) = \frac{\exp \{\bar{x}_i\} }{\sum_{i=1}^{D} \exp \{\bar{x}_i\} }$. {\color{black}Finally, the output layer $f^{(17)}(\cdot)$ is a regression layer of size $L\times 1$.} {\color{black}The current network parameters are obtained from a hyperparameter tuning process providing the best performance for the considered scenario \cite{elbirQuantizedCNN2019,elbirIETRSN2019}. }
	
	The proposed deep networks are realized and trained in MATLAB on a PC with a single GPU and a 768-core processor. We used the stochastic gradient descent algorithm with momentum 0.9  and  updated the network parameters with learning rate $0.01$ and mini-batch size of $128$ samples. Then, we reduced the learning rate by the factor of $0.5$ after each 10 epochs. We also applied a stopping criteria in training so that the training ceases when the validation accuracy does not improve in three consecutive epochs. Algorithm~\ref{alg:algorithmTraining} summarizes the training data generation.

	\section{Numerical Simulations}
	\label{sec:sim}
	In this section, we present the performance of the proposed \textsf{DeepMUSIC} algorithm in comparison with the MLP structure in \cite{deepDOAEstMLP_DoaEstimation}, the  MUSIC algorithm~\cite{music} and the Cramer-Rao lower Bound (CRB)~\cite{crbStoicaNehorai}.  	In the training stage, we use the angular spectrum as $[\Theta^\mathrm{start},\Theta^\mathrm{final}] = [-60^\circ, 60^\circ]$~\cite{deepDOAEstMLP_DoaEstimation}  with $N=2^{12}$ grid points. We select $K=5$ and the target DOAs are randomly located in $Q=8$ subregions. In particular, $J_\alpha$ DOA angle sets $\{\theta_k^{(\alpha)}\}_{k=1}^K$ are realized for $\alpha = 1\dots,J_\alpha$ and the DOA angles are selected as the angles drawn uniform randomly from  $\Theta$. Then, for each realization, noisy array outputs $\mathbf{y}^{(\alpha,\beta)}(t_i)$ are generated for $\beta = 1,\dots, J_\beta$. We select $J_\alpha = J_\beta=100$ and $T = 500$ in out simulations. When generating the data, we use four different SNR levels, i.e., SNR$_\mathrm{TRAIN} = \{ 15, 20, 25, 30\}$ dB where SNR$_\mathrm{TRAIN} = 10\log_{10}(\sigma_\mathrm{S}^2/\sigma_{\mathrm{TRAIN}}^2)$ and  $\sigma_\mathrm{S}^2 = 1$. Hence, the total data set length is $J=4 J_\alpha J_\beta = 40000$.  Further, $80\%$ and $20\%$ of all generated data are chosen for training and validation datasets, respectively. For the prediction process, we select the DOA angles as the floating angles generated uniform randomly in the subregions defined above. Then, $J_T=100$ Monte Carlo experiments are conducted to obtain the statistical performance of the proposed \textsf{DeepMUSIC} framework. Throughout the simulations, we use a ULA with $M=16$ antennas with $\lambda=\bar{d}/2$, $T=100$.

	\begin{figure}[t]
		\centering
		{\includegraphics[draft=false,width=.35\textheight,height=.20\textheight]{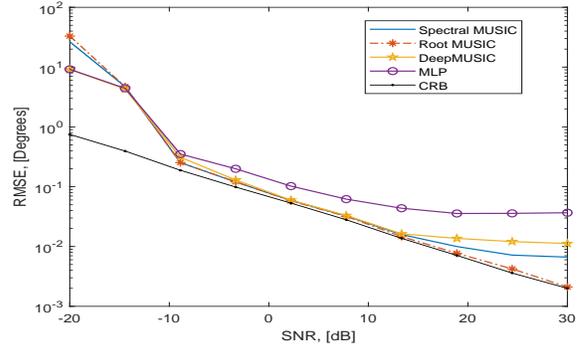} } 
		\caption{DOA estimation performance versus SNR for $K=2$.   }
		\label{fig_RMSE_SNR_K2}
	\end{figure}

	\begin{figure}[t]
		\centering
		{\includegraphics[draft=false,width=.35\textheight,height=.20\textheight]{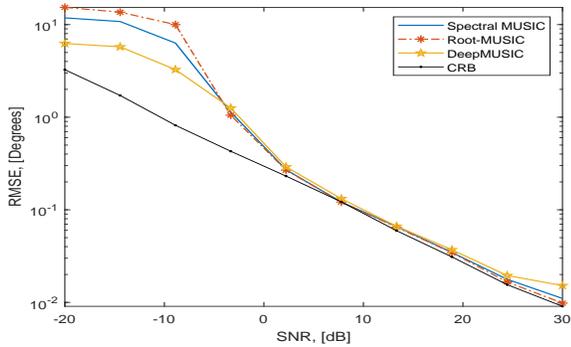} } 	
		\caption{DOA estimation performance versus SNR for  $K=6$.   }
		\label{fig_RMSE_SNR_K6}
	\end{figure}
	
	In Fig.~\ref{fig_RMSE_SNR_K2} and Fig.~\ref{fig_RMSE_SNR_K6} we present the DOA estimation performance of the algorithms for $K=2$ and $K=6$ respectively. When $K=2$, we see that \textsf{DeepMUSIC} outperforms MLP and provides very close performance to the spectral and Root-MUSIC algorithms respectively.  When $K=6$, it can be seen that \textsf{DeepMUSIC} performs better  than the MUSIC algorithms in low SNR regimes and closely  follows the performance of the MUSIC algorithms as SNR increases. The performance of \textsf{DeepMUSIC} can be attributed to the use of convolutional layers which extract the hidden features in the input data whereas the MLP architecture consists of only fully connected layers which do not provide effective feature extraction. Furthermore, MLP only works for two signals which brings a strong limitation for practicality. In contrast, the proposed \textsf{DeepMUSIC} framework can handle multiple target scenario and it exhibits a reasonable performance.
	
	We can also see from Fig.~\ref{fig_RMSE_SNR_K2}-\ref{fig_RMSE_SNR_K6} that, for high SNR regimes, i.e., SNR$\geq 20$ dB, the performance of the DL-based methods and the spectral MUSIC algorithm maxes out and does not improve. One of the main reasons of this is due to reaching the angular resolution limit\footnote{The resolution limit $\gamma$ can also be viewed as the angular search step size of the MUSIC algorithm~\cite{music}.} $\gamma = |-60 - 60|/2^{12} \approx 0.02$, which limits the performance of spectral approaches except Root-MUSIC. Moreover, the performance loss for \textsf{DeepMUSIC} and MLP is due to the loss of precision in the deep networks. This is because, being biased estimators, deep networks do not provide unlimited accuracy. This problem can be mitigated by increasing the number of units in various network layers. Unfortunately, it may lead to the network memorizing the training data and perform poorly when the test data are different than the ones in training. To balance this trade-off, we used noisy data-sets  with several SNR$_\mathrm{TRAIN}$ levels during training so that the network attains reasonable tolerance to corrupted/imperfect inputs.  While similar performance degradation is also observed in~\cite{deepDOAEstSparsePrior,deepDOAestUCA}, no justification is provided for this issue.
	
	\begin{figure}[t]
		\centering
		{\includegraphics[draft=false,width=.35\textheight,height=.20\textheight]{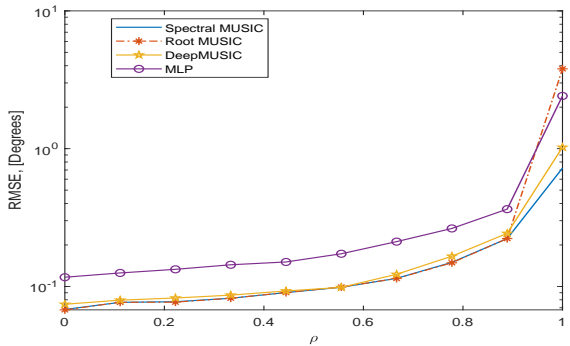} } 
		\caption{DOA estimation performance versus correlation coefficient.   }
		\label{fig_RMSE_correlation}
	\end{figure}
	
	In Fig.~\ref{fig_RMSE_correlation}, the DOA estimation performance is presented with the same simulation settings when there are $K=2$ correlated target signals as $\footnotesize\boldsymbol{\Gamma} = \big[ \begin{array}{cc}
	\sigma_1^2 & \rho \\ \rho & \sigma_2^2 
	\end{array}\big]$ where $\rho$ is the correlation coefficient. We can see that \textsf{DeepMUSIC} closely follows the MUSIC algorithm and provides less RMSE when fully correlated case (i.e., $\rho=1$).
	
	We also compare the computation complexity of the DOA estimation algorithms for the same settings and $K=2$. We observe that \textsf{DeepMUSIC} and MLP take about $0.0020$ s and $0.0110$ s whereas spectral MUSIC and Root-MUSIC need $0.0300$ s and $0.0040$ s to obtain the MUSIC spectra. These results show that \textsf{DeepMUSIC} has the lowest computation time as compared to the competing algorithms. {\color{black}While \textsf{DeepMUSIC} has multiple networks as compared to MLP, it provides less computation time due to the fact that 1) \textsf{DeepMUSIC} has many convolutional layers rather than fully connected layers which involve higher complexity~\cite{vggRef} and 2) the multiple networks in \textsf{DeepMUSIC} can be trained and run with parallel processing so that the computational complexity is reduced. Similar observations are also reported in~\cite{deepDOAEstSparsePrior}. }


	%
	
	%
	%
	%

	\section{Summary}
	In this letter, we introduced a DL framework called \textsf{DeepMUSIC} for DOA estimation. The major advantage of the  proposed approach is that it can work for multiple targets in comparison with the previous works. Furthermore, \textsf{DeepMUSIC} provides less computational complexity as compared to the conventional techniques.

	\bibliographystyle{IEEEtran}
	\footnotesize{\bibliography{IEEEabrv,references_052_journal}}
	\balance

\end{document}